\documentclass[conference]{IEEEtran}

\ifCLASSINFOpdf
\usepackage[pdftex]{graphicx}

\else

\fi

\usepackage[cmex10]{amsmath}
\usepackage{amssymb}   % \triangleq 
\usepackage{amsxtra}
\usepackage{amscd}
\usepackage{amsthm}
\usepackage{textcomp}
\usepackage{graphicx}  
\usepackage{balance}
\usepackage{array}  
\usepackage{multirow}  
\usepackage{amsmath}
\usepackage{cite}
\usepackage{times}
\usepackage{blindtext}

\newcommand\blfootnote[1]{%
	\begingroup
	\renewcommand\thefootnote{}\footnote{#1}%
	\addtocounter{footnote}{-1}%
	\endgroup}

\usepackage{url}
\usepackage{color,soul} % for highlights
\soulregister\cite7
\soulregister\ref7
\soulregister\pageref7
\begin{document}
	
	\title{{\huge SimRIS Channel Simulator for Reconfigurable Intelligent Surface-Empowered Communication Systems}}

	\author{\IEEEauthorblockN{Ertugrul Basar\textsuperscript{$\ast$} and Ibrahim Yildirim\textsuperscript{$\ast$,$\bullet$ }}
		\IEEEauthorblockA{\textsuperscript{$\ast$}CoreLab, Department of Electrical and Electronics Engineering, Ko\c{c} University, Sariyer 34450, Istanbul, Turkey  \\
			\textsuperscript{$\bullet $}Faculty of Electrical and Electronics Engineering, Istanbul Technical University, Sariyer  34469, Istanbul, Turkey.  \\
			Email: ebasar@ku.edu.tr, yildirimib@itu.edu.tr}}

	\maketitle

	\begin{abstract}
		Reconfigurable intelligent surface (RIS)-assisted communication appears as one of the potential enablers for sixth generation (6G) wireless networks by providing  a new way to optimize the communication system performance. This paper aims to fill an important gap in the open literature by providing an accurate, open-source, and widely applicable RIS channel model for mmWave frequencies. Our model is not only applicable in various indoor and outdoor environments but also includes the physical aspects of wireless propagation in the presence of an RIS as well as various practical 5G channel modeling issues. The open-source and comprehensive \textit{SimRIS Channel Simulator} is also introduced in this paper to be used in computer simulations of RIS-assisted communication systems.
		
	\end{abstract}
	
	\begin{IEEEkeywords}
		6G, channel modeling, millimeter wave, reconfigurable intelligent surface (RIS).
	\end{IEEEkeywords}

	%	\linenumbers

	\IEEEpeerreviewmaketitle
	
	%\vspace*{-0.22cm}
	\section{Introduction}

	\IEEEPARstart{S}{ixth} generation (6G) wireless systems are expected to provide broadband and ubiquitous connectivity by supporting new use-cases including extreme capacity and very high mobility, integrated with satellite networks and autonomous systems \cite{Rajatheva_6G}. These attractive features can be supported by utilizing effective physical layer (PHY) tools such as ultra massive multiple-input multiple-output (MIMO) systems, millimeter wave (mmWave) and TeraHertz (THz) communications, and reconfigurable intelligent surfaces (RISs). \blfootnote{This work was supported in part by TUBITAK under Grant 117E869.
		
		MATLAB package and user interface of \textit{SimRIS Channel Simulator} is available at  https://corelab.ku.edu.tr/tools/SimRIS}

	RIS-empowered communication has received growing interest from the wireless research community due to its undeniable potential in extending the coverage, enhancing the link capacity, mitigating interference, deep fading, and Doppler effects, and increasing the PHY security \cite{Akyildiz_2019,Wu_2019,Basar_Access_2019,Basar_Doppler,yildirim2019propagation}. RISs enable the control of wireless propagation through their unique electromagnetic functionalities and provide a new degree of freedom in the system design.

RIS-assisted mmWave communication systems have recently attracted the attention of the research community and promising results have been reported in this context  \cite{Hey_2020,Yang_2020}. In \cite{Hey_2020}, an RIS in the form of a uniform linear array (ULA) is considered, which might be difficult to implement in practice, while \cite{Yang_2020} deals with point-to-point mmWave links only with ULA-type RISs.  The physical aspects and modeling of RIS-empowered communication systems have also been broadly investigated in the past few months by researchers. While useful insights for the performance limits of RIS-empowered systems are provided in \cite{Basar_Access_2019} by considering specular reflection, path loss effects are not taken into account. Although \cite{Ellingson} introduces a signal model by considering plate scattering and radar range paradigms along with the physical area and practical gain of RIS elements, this model might only be useful for a static environment. 
		
Against this background, there is an urgent need for a physical and widely applicable mmWave channel model to be used in various RIS-assisted systems in indoor and outdoor environments. 	Considering that RIS channel modeling is the first step towards RIS-empowered networks, this paper aims to provide a new line of research by integrating RISs into state-of-the-art 5G channel models. Within this context, our major contributions are summarized below:
	\begin{itemize}
		\item We put forward fundamental channel models and power scaling laws for RIS-assisted systems in the presence of multiple scatterers and formulate a baseline cascaded channel model.

		\item We introduce the open-source \textit{SimRIS Channel Simulator} MATLAB package, which can be used in channel modeling and computer simulations of RIS-based communication systems with tunable operating frequency, terminal and RIS locations, number of RIS elements, and environments.
		\item Using SimRIS Channel Simulator, we provide preliminarily computer simulation results and reveal the practical use-cases of RISs for indoors and outdoors. 
	\end{itemize}
	
	The rest of the paper is organized as follows. In Section II, we present the fundamentals and introduce our baseline RIS-based channel models. In Section III, we introduce our RIS-assisted channel model for indoors and outdoors. In Section IV, \textit{SimRIS Channel Simulator} is presented, while Section V presents our numerical results. Finally, the paper is concluded in Section VI. \vspace*{-0.3cm}

\section{RIS-Assisted Communications: Fundamentals}
In this section, we revisit the link power equation and the channel model of an RIS-assisted system without any obstacles or reflecting/scattering elements (interacting objects, IOs) between the terminals and the RIS. Then we extend our model considering the existence of multiple IOs between the Tx and the RIS to model a more practical scenario.

\begin{figure}[!t]
	\begin{center}
		\includegraphics[width=0.8\columnwidth]{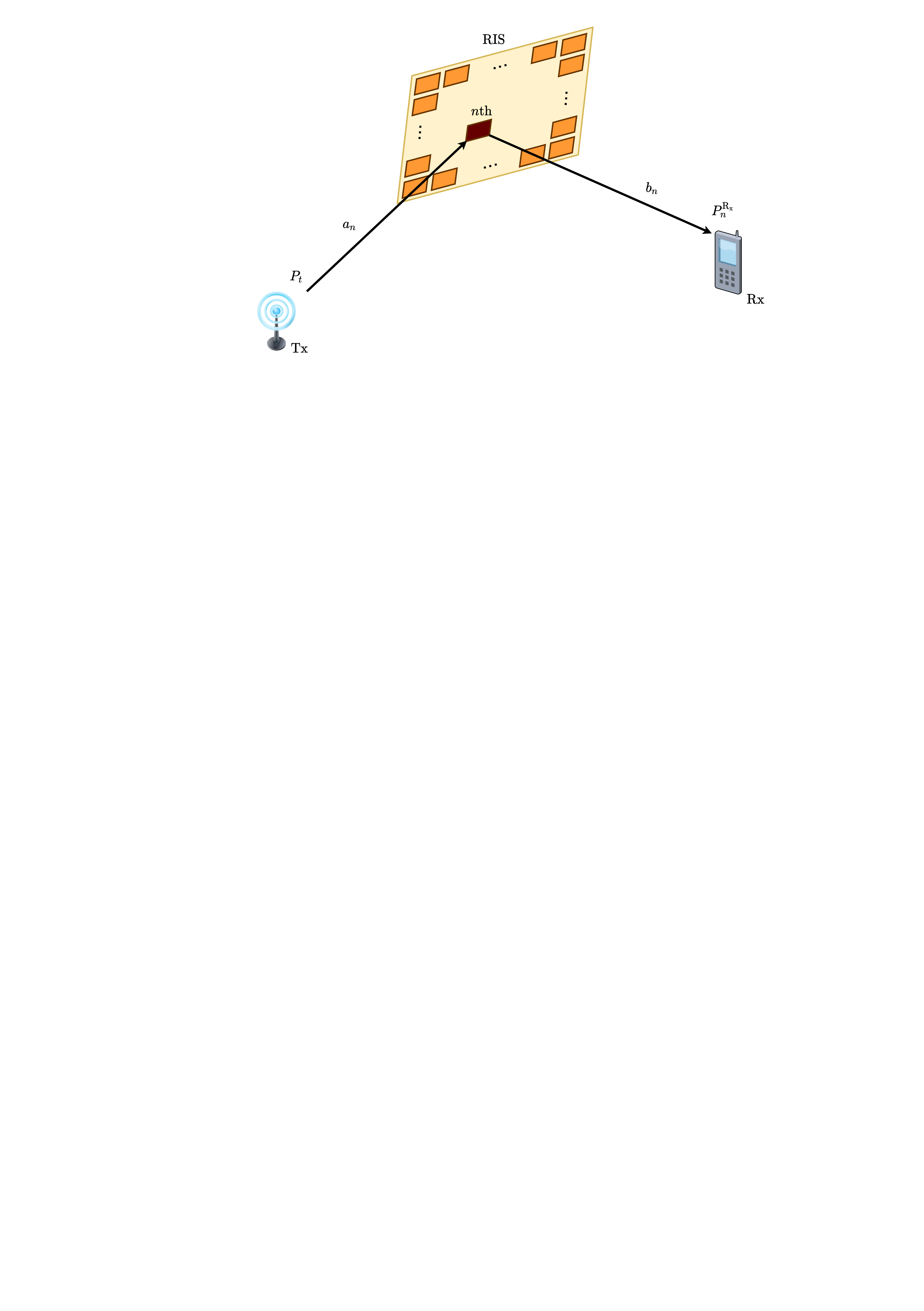}
		\vspace*{-0.3cm}\caption{RIS-assisted communication with LOS links.}\vspace*{-0.2cm}
		\label{fig:Fig1}
	\end{center}\vspace*{-0.5cm}
\end{figure}
\subsection{RIS-Assisted LOS Channels}
The considered fundamental RIS-based communication system is shown in Fig. 1 with pure LOS links between terminals. According to the plate scattering theory, the transmitted signal is captured by each RIS element, then re-scattered to the medium in all directions. Focusing on the $n$th RIS element, the captured power on it can be readily obtained as
\begin{equation}\label{eq:1}
P_n^{\text{RIS }}=\dfrac{P_t G_t G_{e}^{\text{Tx}} \lambda^2}{(4 \pi)^2 a_n^2}
\end{equation}
where $P_t$ is the transmit power, $G_t$ is the transmit antenna gain in the direction of the $n$th RIS element (or the RIS in general), $G_{e}^{\text{Tx}}$ is the gain of the corresponding RIS element in the direction of the transmitter (Tx), $\lambda $ is the wavelength, and $a_n$ is the distance between the transmitter and this element. As seen from \eqref{eq:1}, considering the effective aperture of the $n$th RIS element given by $ G_{e}^{\text{Tx}} \lambda^2 /(4\pi)$ and the power flux density incident on it given by $P_t G_t /(4\pi a_n^2)$, the captured power at the RIS element follows the free-space propagation as expected. Considering the passive nature of RIS elements, the captured power is re-radiated to the medium with an efficiency factor $\epsilon$, which is assumed to be unity (for maximum efficiency). In light of this information, denoting the distance between this element and the receiver by $b_n$, the captured power at the receiver (Rx) is obtained as
\begin{equation}\label{eq:2}
P_{n}^{\text{Rx }}=\dfrac{P_n^{\text{RIS }} G_{e}^{\text{Rx}} G_r \lambda^2 }{ (4\pi)^2 b_n^2 }= \dfrac{P_t G_t G_r G_{e}^{\text{Tx}} G_{e}^{\text{Rx}}\lambda^4}{(4\pi)^4 a_n^2 b_n^2}
\end{equation}
where $G_r$ is the receive antenna gain in the direction of the $n$th RIS element and $G_{e}^{\text{Rx}}$ is the gain of the corresponding RIS element in the direction of the receiver. For the same scenario, let us consider the radar range equation given by 
\begin{equation}\label{eq:3}
P_r=\frac{P_t G_t G_r\lambda^2 \sigma_{\text{RCS}}}{(4\pi)^3 a_n^2 b_n^2} 
\end{equation}
where $\sigma_{\text{RCS}}=4 \pi A^2 / \lambda^ 2$ is the radar cross section (RCS, in $\text{m}^2$) \cite{Stutzman} of the RIS element with $A$ being its physical area. One can easily notice the conceptual similarity between two models assuming an isotropic scatterer with $G_e=G_{e}^{\text{Tx}}=G_{e}^{\text{Rx}}=4 \pi A_e/\lambda^2$, where $A_e \sim A$ is the effective aperture of the RIS element. Plugging this in the radar cross section formula given above yields $\sigma_{\text{RCS}}=\lambda^ 2 G_e^2 / 4\pi  $, which gives the same result in \eqref{eq:2} when applied in \eqref{eq:3}.

For clarity of presentation, let us assume $P_t=0$ dBW and isotropic antennas at the Tx and Rx  terminals by $G_t=G_r=1\, (0\text{ dBi})$. We further assume that  $G_e=G_{e}^{\text{Tx}}=G_{e}^{\text{Rx}}$, which might hold when the angles of incidence and departure are equal to each other. It is worth noting that the gain of RIS elements depends on the angles of incidence and departure, however, $G_e=\pi$ ($5$ dBi) can be taken as a reference as the maximum element gain for the broadside incidence \cite{Ellingson}, while a generalized element radiation pattern is considered in the SimRIS Channel Simulator. In light of these, we obtain
\begin{equation}\label{eq:4}
P_{n}^{\text{Rx }}= \dfrac{ G_{e}^2\lambda^4}{(4\pi)^4 a_n^2 b_n^2}.
\end{equation}
Due to the scattering nature of RIS elements, we can state that the total path attenuation (gain) $(L_n=P_{n}^{\text{Rx }}/P_t)$ is the product of the individual path attenuations of two links $(L_{n,1} \text{ and } L_{n,2})$, that is
\begin{equation}\label{eq:5}
L_n= L_{n,1} \times L_{n,2} = \dfrac{G_{e} \lambda^2 }{(4\pi)^2 a_n^2} \times \dfrac{G_{e} \lambda^2 }{(4\pi)^2 b_n^2}
\end{equation} 
which is one of the fundamental properties of RIS-assisted communication links and will be used later.

\begin{figure}[!t]
	\begin{center}
		\includegraphics[width=0.9\columnwidth]{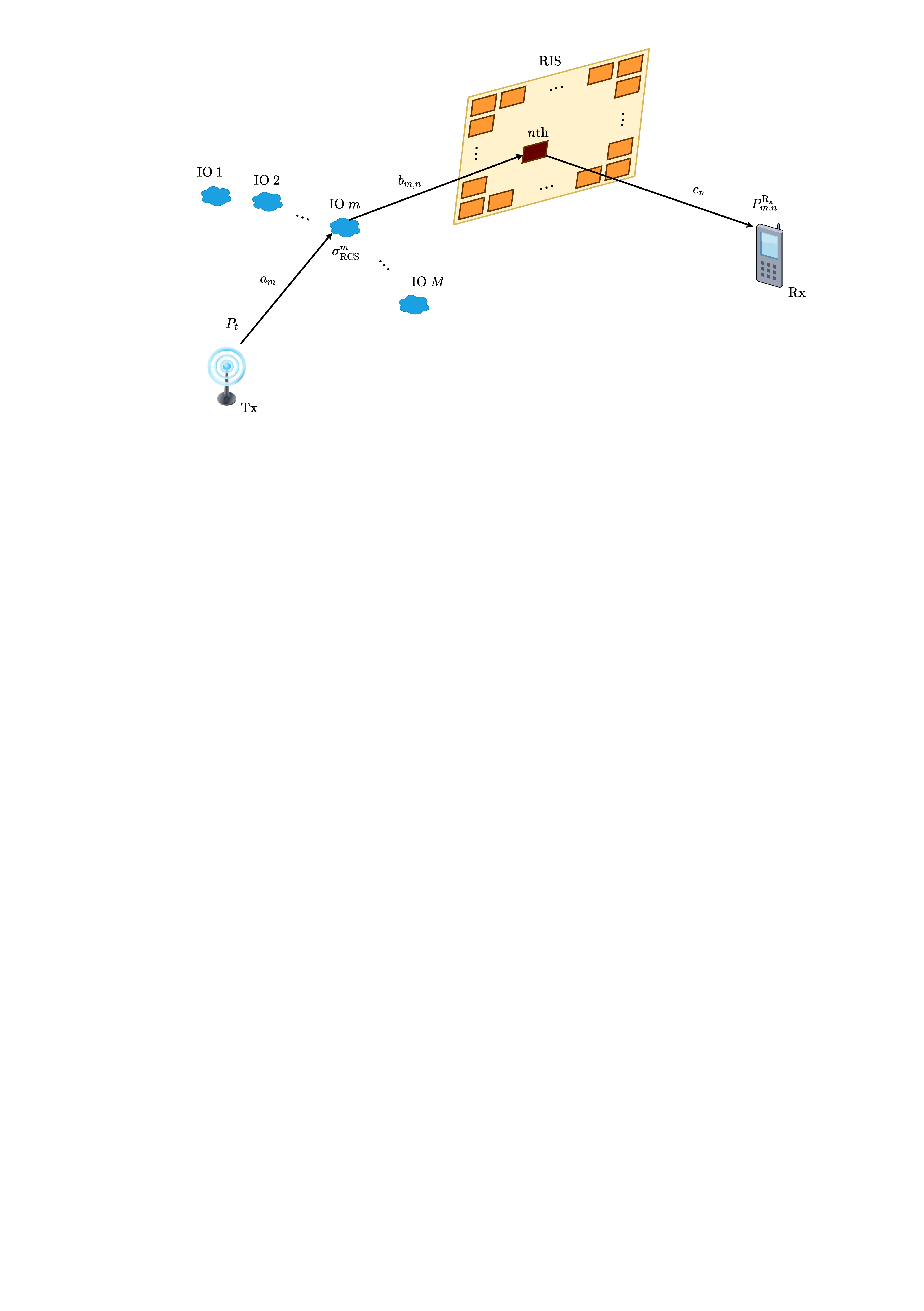}
		\vspace*{-0.3cm}\caption{RIS-assisted communication with $M$ IOs between Tx-RIS.}\vspace*{-0.4cm}
		\label{fig:Fig2}
	\end{center} \vspace*{-0.2cm}
\end{figure}

Generalizing the scattering concept given above for an RIS with $N$ elements, the received discrete-time baseband (noise-free) signal can be expressed as
\begin{equation}\label{eq:6}
y= \left( \sum\nolimits_{n=1}^{N} \sqrt{P_{n}^{\text{Rx}}} ( \alpha_n e^{j\phi_n})  e^{-j k (a_n+b_n)} \right) x 
\end{equation} 
where $\alpha_n$ and $\phi_n$ respectively stand for controllable magnitude and phase response of the $n$th element, $k=2\pi/\lambda$ is the wave number, and $x$ is the transmitted signal. One can easily observe from \eqref{eq:6} that the received signal power can be maximized by adjusting RIS element phases as $\phi_n= k(a_n+b_n)$, i.e., by aligning the phases of incoming signals. Finally, it is worth noting that the direct link between Tx and Rx can be incorporated into the model by
\begin{equation}\label{eq:6_ek}
y= \left( \sum\limits_{n=1}^{N} \sqrt{P_{n}^{\text{Rx}}} ( \alpha_n e^{j\phi_n})  e^{-j k (a_n+b_n)} + \sqrt{P_{\text{T-R}}} e^{-jkd_{\text{T-R}}} \right) x 
\end{equation} 
where $P_{\text{T-R}} = \lambda^2 /(4\pi d_{\text{T-R}} )^2$ is the received LOS power with $d_{\text{T-R}}$ being the Tx-Rx distance.

Although providing a baseline, the model shown above is not applicable to practical RIS-assisted communication scenarios due to its dependency on pure LOS links between the RIS and Tx/Rx terminals.

\subsection{RIS-Assisted Channels with Interacting Objects}
In this subsection, we relax the system model of previous section by assuming $M$ IOs (scatterers) between the transmitter and the RIS as shown in Fig. 2, while assuming a pure LOS link between the RIS and the Rx. Here, $a_m$, $b_{m,n}$, and $c_n$ respectively stand for the distances between Tx and the $m$th IO, the $m$th IO and the $n$th RIS element, and the $n$th RIS element and Rx. Furthermore, RCS of the $m$th IO is shown by $\sigma_{\text{RCS}}^m$. Focusing on the particular scenario of one IO and one RIS element, according to the radar range equation of \eqref{eq:3}, the captured power at the $n$th RIS element is given by
\begin{equation}\label{eq:7}
P_{m,n}^{\text{RIS }}=\frac{G_e\lambda^2 \sigma_{\text{RCS}}^m}{(4\pi)^3 a_m^2 b_{m,n}^2}. 
\end{equation}
Applying the same procedures in \eqref{eq:2}, the received power at the receiver can be obtained as
\begin{equation}\label{eq:8}
P_{m,n}^{\text{Rx}} = \dfrac{P_{m,n}^{\text{RIS }} G_{e} \lambda^2 }{ (4\pi)^2 c_n^2 } = \dfrac{G_e^2 \lambda^4 \sigma_{\text{RCS}}^m}{(4\pi)^5 a_m^2 b_{m,n}^2 c_n^2}.
\end{equation}
As seen from \eqref{eq:8}, the overall path gain (attenuation) is again obtained as the product of the path gains of individual paths, which are related by $1/a_m^2$, $1/b_{m,n}^2$, and $1/c_n^2$, respectively. As a result, the received signal is given by
\begin{equation}\label{eq:9}
y= \left(  \sqrt{P_{m,n}^{\text{Rx}}}  (\alpha_n e^{j\phi_n})  e^{-j k \psi_{m,n}} \right) x 
\end{equation} 
where $\psi_{m,n}=a_m+b_{m,n}+c_n $. Let us generalize this model to case of $M$ IOs and $N$ RIS elements, where the received signal, similar to double scattering scenario of \cite{Andersen_2002}, is obtained as
\begin{equation}\label{eq:10}
y= \left( \sum\nolimits_{n=1}^{N}  \alpha_n e^{j\phi_n} \left(  \sum\nolimits_{m=1}^{M}  \sqrt{P_{m,n}^{\text{Rx}}}    e^{-j k \psi_{m,n}} \right) \right) x .
\end{equation} 
Here, each RIS element collects all signals scattered from $M$ IOs and forwards to the receiver, where the received signal is the sum of $N$ scattered components.

\begin{figure*}[!t]
	\begin{center}
		\includegraphics[width=1.7\columnwidth]{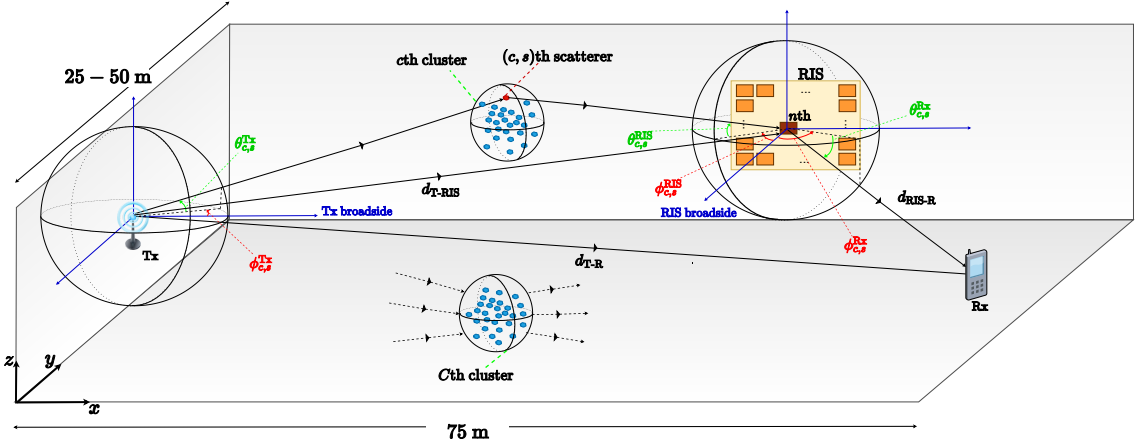}
		\vspace*{-0.3cm}\caption{Generic InH Indoor Office environment with $C$ clusters between Tx-RIS and an RIS mounted in the $xz$ plane (side wall).}\vspace*{-0.7cm}
		\label{fig:Fig4}
	\end{center}
\end{figure*}
Let us re-express \eqref{eq:10} by separating the channel gains  belonging to  Tx-RIS and RIS-Rx links:
\begin{align}\label{eq:11}
y&= \left( \sum\limits_{n=1}^{N} \sqrt{L_n^{\text{LOS}}}  (\alpha_n e^{j\phi_n} )e^{-jkc_n} \nonumber \right. \\
&\hspace*{1cm}\left. \times  \left(  \sum\limits_{m=1}^{M}  \sqrt{L_{m,n}^{\text{RIS}}}    e^{-j k (a_m+b_{m,n})} \right) \right) x 
\end{align}
where $L_n^{\text{LOS}}=G_e\lambda^2  / (4\pi c_n)^2$ and $L_{m,n}^{\text{RIS}}=G_e \lambda^2 \sigma_{\text{RCS}}^m/((4\pi)^3 a_m^2 b_{m,n}^2)$ stand for the path gains of the LOS (RIS-Rx) path and the Tx-IOs-RIS path, and we have $P_{m,n}^{\text{Rx}}=L_n^{\text{LOS}} L_{m,n}^{\text{RIS}}$. 
\eqref{eq:11} can be rewritten in vector form as
\begin{equation}\label{eq:12}
y= \mathbf{g}^{\mathrm{T}} \mathbf{\Theta} \mathbf{h}x 
\end{equation}
where $\mathbf{g} = \begin{bmatrix}
\sqrt{L_{1}^{\text{LOS}} }  e^{-jkc_1 }& \cdots & \sqrt{L_N^{\text{LOS}} } e^{-jkc_N }
\end{bmatrix}^{\mathrm{T}}$ is the vector of LOS channel coefficients between the RIS and Rx, $ \mathbf{\Theta} = \mathrm{diag} ( \begin{bmatrix}
\alpha_1 e^{j\phi_1} & \cdots & \alpha_N e^{j\phi_N} 
\end{bmatrix}   ) $ is the matrix of RIS element responses, and $\mathbf{h}=\begin{bmatrix}
\sum\limits_{m=1}^{M}\sqrt{L_{m,1}^{\text{RIS}}} e^{-j k (a_m+b_{m,1})} & \cdots & \sum\limits_{m=1}^{M}\sqrt{L_{m,N}^{\text{RIS}}} e^{-j k (a_m+b_{m,N})}
\end{bmatrix}$
is the vector of channel coefficients for the Tx-RIS link composed of $M$ scatterers. It is worth noting that when the RIS is far from the Tx and the Rx, $b_{m,n}$ and $c_n$ may be assumed to be independent of the RIS element $n$ from the perspective of channel gains, but not from the phases, and \eqref{eq:12} can be further simplified. Similar to \eqref{eq:6_ek}, the direct signal component between Tx and Rx (with or without IOs) can also be included in \eqref{eq:12} as \vspace*{-0.1cm}
\begin{equation}\label{eq:LOS}
y= \left( \mathbf{g}^{\mathrm{T}} \mathbf{\Theta} \mathbf{h} + h_{\text{SISO}} \right) x 
\end{equation}
where $h_{\text{SISO}}$ characterizes the direct link (narrowband) channel between Tx and Rx, which is equal to $h_{\text{SISO}}=\sqrt{P_{\text{T-R}}} e^{-jkd_{\text{T-R}}}$ for a LOS dominated link.
	
Before concluding this section, we note that \eqref{eq:10} can be used to model the power budget of an RIS-assisted system operating in a more realistic environment with multiple IOs. However, this model again does not capture the variations of the environmental objects and transmit/receiver movements, i.e., based on a static setup. \vspace*{-0.1cm}

\section{Practical Channel Modeling: Indoors and Outdoors}
In this section, we introduce an unified signal/channel model for RIS-assisted 6G communication systems operating under mmWave frequencies. This model is generic and can be applied to different environments (indoors/outdoors) and operating frequencies. Considering the promising potential of RIS-assisted systems for mmWave communications, we build our framework on the clustered statistical MIMO model, which is widely used in 3GPP standardization \cite{3GPP_5G}, while a generalization is possible. 	Due to their unique functionality, channel modeling methedology for RISs is different from relay technologies. Although the major steps of RIS-assisted channel modeling are provided in this paper, interested readers are referred to \cite{SimRIS_jour_new} for the technical details and statistical background of the considered channel parameters. In the following, we summarize our methodology to generate Tx-RIS, RIS-Rx and Tx-Rx subchannels for indoors and outdoors. Here, we consider that the Tx lies on the $ yz $ plane, while the RIS lies either on the $ xz $ plane (Scenario 1 - side wall) or $ yz $ plane (Scenario 2 - opposite wall) for indoor and outdoor environments. The considered 3D geometry for a typical large indoor office is given in Fig. \ref{fig:Fig4} as a reference for Scenario 1. This 3D geometry can be easily extended to outdoor environments by modifying certain system parameters.
%\subsection{mmWave Indoor Channel Modeling}

We assume that the existing IOs are groped under $C$ clusters, each having $S_c$ sub-rays for $c=1,\ldots,C$, that is $M=\sum_{c=1}^{C}S_c$. Therefore, the vector of Tx-RIS channel coefficients $\mathbf{h} \in \mathbb{C}^{N\times 1}$ in \eqref{eq:12} and \eqref{eq:LOS} can be rewritten for a clustered model by considering array responses and path attenuations:
\begin{equation}\label{eq:13}
\mathbf{h}=
\gamma \sum\limits_{c=1}^{C} \sum\limits_{s=1}^{S_c} \beta_{c,s} \sqrt{G_e(\theta_{c,s}^{\text{RIS}})L_{c,s}^{\text{RIS}}} \,\, \mathbf{a} ( \phi_{c,s}^{\text{RIS}}, \theta_{c,s}^{\text{RIS}} ) + \mathbf{h}_{\text{LOS}}
\end{equation}
where $ \gamma= \sqrt{\frac{1}{\sum\nolimits_{c=1}^{C} S_c}}$ is a normalization factor, $\mathbf{h}_{\text{LOS}}$ is the LOS component, $\beta_{c,s} \sim \mathcal{CN} (0,1)$ and $L_{c,s}^{\text{RIS}}$ respectively stand for the complex path gain and attenuation associated with the $(c,s)$th propagation path, and $G_e(\theta_{c,s}^{\text{RIS}})$ is the RIS element pattern \cite{Nayeri} in the direction of the $(c,s)$th scatterer. Here,  $\mathbf{a} ( \phi_{c,s}^{\text{RIS}}, \theta_{c,s}^{\text{RIS}} ) \in \mathbb{C}^{N\times 1}$ is the array response vector of the RIS for the considered azimuth ($ \phi_{c,s}^{\text{RIS}}$) and elevation 
($\theta_{c,s}^{\text{RIS}}$) arrival angles (with respect to the RIS broadside) and carefully calculated for our system due to the fixed orientation of the RIS. 
Here, the number of clusters, number of sub-rays per cluster, and the locations of the clusters can be determined for a given environment and frequency.

For the attenuation of the $(c,s)$th path, we adopt the 5G path loss model (the
close-in free space reference distance model with frequency-dependent path loss exponent, in dB), which is applicable to various environments including Urban Microcellular (UMi) and Indoor Hotspot (InH) \cite{5G_Channel}.
The LOS component of $\mathbf{h}$ is calculated by
\begin{equation}\label{eq:hLOS}
\mathbf{h}_{\text{LOS}}= I_{\mathbf{h}}(d_{\text{T-RIS}}) \sqrt{G_e(\theta_{\text{LOS}}^{\text{RIS}}) L_{\text{LOS}}^{\text{T-RIS}}}  e^{j\eta} \mathbf{a}(\phi_{\text{LOS}}^{\text{RIS}},\theta_{\text{LOS}}^{\text{RIS}})
\end{equation}
where $L_{\text{LOS}}^{\text{T-RIS}}$ is the attenuation of the LOS link, $G_e(\theta_{\text{LOS}}^{\text{RIS}})$ is the RIS element gain in the LOS direction, $\mathbf{a}(\phi_{\text{LOS}}^{\text{RIS}},\theta_{\text{LOS}}^{\text{RIS}})$ is the array response of the RIS in the direction of the Tx, and $\eta \sim \mathcal{U} [0,2\pi]$. Here, $I_{\mathbf{h}}(d_{\text{T-RIS}})$ is a Bernoulli random variable taking values from the set $\left\lbrace 0,1 \right\rbrace $ and characterizes the existence of a LOS link for a Tx-RIS separation of $d_{\text{T-RIS}}$. It is again calculated according to the 5G model \cite{5G_Channel}.

For the calculation of LOS-dominated RIS-Rx channel $\mathbf{g}$ in an indoor environment, we re-calculate the RIS array response in the direction of the Rx by calculating azimuth and elevation departure angles $\phi^{\text{RIS}}_{\text{Rx}}$ and $\theta^{\text{RIS}}_{\text{Rx}}$ for the RIS from the coordinates of the RIS and the Rx. Finally,  the vector of LOS channel coefficients can be generated as
\begin{equation}\label{eq:20}
\mathbf{g}=\sqrt{G_e(\theta^{\text{RIS}}_{\text{Rx}}) L_{\text{LOS}}^{\text{RIS-R}}} e^{j\eta} \mathbf{a}(\phi^{\text{RIS}}_{\text{Rx}},\theta^{\text{RIS}}_{\text{Rx}}).
\end{equation}
where $G_e(\theta^{\text{RIS}}_{\text{Rx}})$ is the gain of RIS element in the direction of the Rx, $L_{\text{LOS}}^{\text{RIS-R}}$ is the attenuation of LOS RIS-Rx channel, $\eta \sim \mathcal{U} [0,2\pi]$ is the random phase term and $\mathbf{a}(\phi^{\text{RIS}}_{\text{Rx}},\theta^{\text{RIS}}_{\text{Rx}})$ is the RIS array response in the direction of the Rx.

For outdoor channel modeling, the major change will be in the channel between the RIS and the Rx, which might be subject to small-scale fading as well with a random number of unique clusters. For this case, we have
\begin{equation}\label{eq:outdoor}
\mathbf{g}=
\bar{\gamma} \sum\limits_{c=1}^{\bar{C}} \sum\limits_{s=1}^{\bar{S_c}} \bar{\beta}_{c,s} \sqrt{G_e(\theta_{c,s}^{\text{Rx}}) L_{c,s}^{\text{Rx}}} \,\, \mathbf{a} ( \phi_{c,s}^{\text{Rx}}, \theta_{c,s}^{\text{Rx}} ) + \mathbf{g}_{\text{LOS}}
\end{equation}
where, similar to \eqref{eq:13}, $\bar{\gamma}$ is a normalization term, $\bar{C}$ and $\bar{S_c}$ stand for number of clusters and sub-rays per cluster for the RIS-Rx link, $\bar{\beta}_{c,s}$ is the complex path gain, $ L_{c,s}^{\text{Rx}} $ is the path attenuation, $G_e(\theta_{c,s}^{\text{Rx}})$ is the RIS element radiation pattern in the direction of the $(c,s)$th scatterer, $\mathbf{a} ( \phi_{c,s}^{\text{Rx}}, \theta_{c,s}^{\text{Rx}} )$ is the array response vector of the RIS for the given azimuth and elevation angles, and $\mathbf{g}_{\text{LOS}}$ is the LOS component. 

The RIS-assisted channel has a double-scattering nature, as a result, the single-scattering link between the Tx and Rx has to be taken into account in our channel model. Even if the RIS is placed near the Rx, the Tx-Rx channel is relatively stronger than the RIS-assisted path, and cannot be ignored in the channel model. 

For indoors, using single-input single-output (SISO) mmWave channel modeling, the channel between these two terminals can be easily obtained (by ignoring arrival and departure angles) as
\begin{equation}\label{eq:18}
h_{\text{SISO}}=
\gamma \sum\limits_{c=1}^{C} \sum\limits_{s=1}^{S_c} \beta_{c,s} e^{j\eta_e} \sqrt{L_{c,s}^{\text{SISO}}} + h_{\text{LOS}}
\end{equation}
where $\gamma$, $C$, $S_c$, and $\beta_{c,s}$ are as defined in \eqref{eq:13} and remain the same for the Tx-Rx channel under the assumption of shared clusters with the Tx-RIS channel, while $h_{\text{LOS}}$ is the LOS component. Here, $L_{c,s}^{\text{SISO}}$ stands for the path attenuation for the corresponding link and $\eta_e$ is the excess phase caused by different travel distances of Tx-RIS and Tx-Rx links over the same scatterers. 

For outdoor environments, we assume that the RIS and the Rx are not too close to ensure that they have independent clusters (small scale parameters) as in the 3GPP 3D channel model \cite{3GPP_5G}. Using SISO mmWave channel modeling, the Tx-Rx channel can be easily obtained as
\begin{equation}\label{eq:SISO}
h_{\text{SISO}}=
\tilde{\gamma} \sum\limits_{c=1}^{\tilde{C}} \sum\limits_{s=1}^{\tilde{S}_c} \tilde{\beta}_{c,s} \sqrt{L_{c,s}^{\text{SISO}}} + h_{\text{LOS}}
\end{equation}
where the number of clusters $\tilde{C}$, sub-rays per cluster $\tilde{S}_c$, complex path gain $\tilde{\beta}_{c,s}$, and path attenuation $L_{c,s}^{\text{SISO}}$ are determined as discussed earlier for the Tx-RIS path and $ \tilde{\gamma}$ is the normalization term. 
%Similar to the clusters/scatterers in Tx-RIS channel, considering the LOS distance $d_{\text{T-R}}$ between the Tx and the Rx, a number of clusters are randomly generated with their conditional Laplacian distributed departure/arrival angles. However, only the total link distances, shown by $\tilde{d}_{c,s}$, are calculated to obtain $L_{c,s}^{\text{SISO}}$ for all $c$ and $s$ during the generation of $h_{\text{SISO}}$. 

\begin{figure}[!t]
	\begin{center}
		\includegraphics[width=0.75\columnwidth]{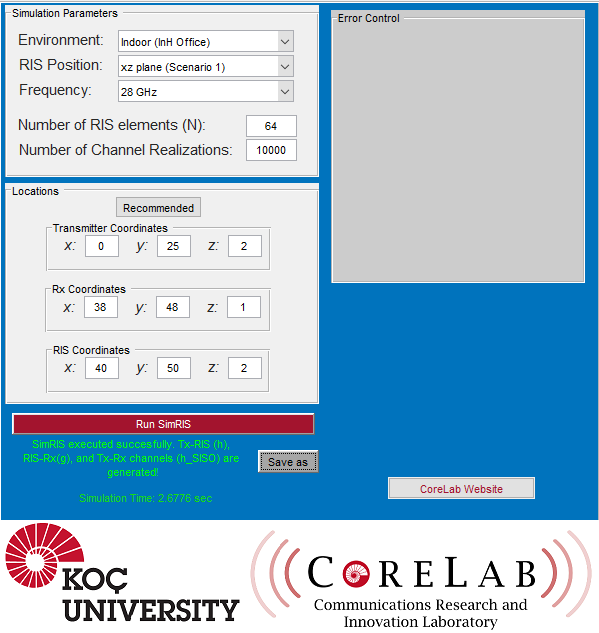}
		\vspace*{-0.1cm}\caption{ Graphical user interface of the SimRIS Channel Simulator v1.0.}\vspace*{-0.5cm}
		\label{fig:GUI}
	\end{center}
\end{figure}

\section{Introduction to  {\normalfont SimRIS} Channel Simulator}

In this section, we introduce the open-source and MATLAB-based\textit{ SimRIS Channel Simulator},  which can be used in channel modeling of RIS-assisted systems with tunable operating frequency, terminal locations, number of RIS elements, and environments. The graphical user interface (GUI) of our \textit{SimRIS Channel Simulator} is given in Fig. \ref{fig:GUI}. For InH Indoor Office and UMi Street Canyon environments, $ \mathbf{h}$, $\mathbf{g}$ and $h_{\text{SISO}}$ channels can be produced by performing Monte Carlo simulations at $28$ and $73$ GHz frequencies by following the procedures briefly described in Section III. \textit{SimRIS Channel Simulator} considers the scattering nature of RIS elements, their physical gains, and unique array responses due to fixed orientation of RISs. Furthermore, it follows a new methodology to calculate the LOS probabilities for the corresponding links by considering the height of the RIS. 

Number of RIS elements ($N$) and number of channel realizations are user-selectable input parameters of the \textit{SimRIS Channel Simulator} as well as the Tx, Rx and RIS locations by considering our 3D geometry given in Fig. \ref{fig:Fig4}. Furthermore, the RIS position determines the selected scenario and can be selected as $xz$ plane (side wall) or $yz$ plane (opposite wall) for both environments in our GUI. Side or opposite wall positions of the RIS respectively represent Scenarios 1 and 2. According to the selected scenario, it is necessary to ensure the consistency of the Tx, the Rx, and the RIS locations. Considering this, with the help of the "Recommendation" button, an example set of positions suitable for the selected scenario and environment is presented to the user for initial operation, while they can be adjusted freely.

 While the channels with the specified number of realizations are generated to the users for their own experiments, warning messages are given about problematic parameters with the error control window. Specifically, the typical cell radius for InH Indoor Office and UMi Street Canyon should be less than $ 75 $ m and $ 100 $ m, respectively. Moreover, the Tx can be mounted at a height of $ 2 $-$ 3 $ m in InH Indoor Office, while $ 3 $-$ 20 $ m in UMi Street Canyon environment. For both cases, the Rx can be deployed as a ground user and its height should be less than $ 2 $ m. The error control box checks whether these and many more practical conditions are met, and gives corresponding warnings and useful instructions to the users.

\section{Numerical Results}
\begin{figure}[!t]
	\begin{center}
		\includegraphics[width=0.9\columnwidth]{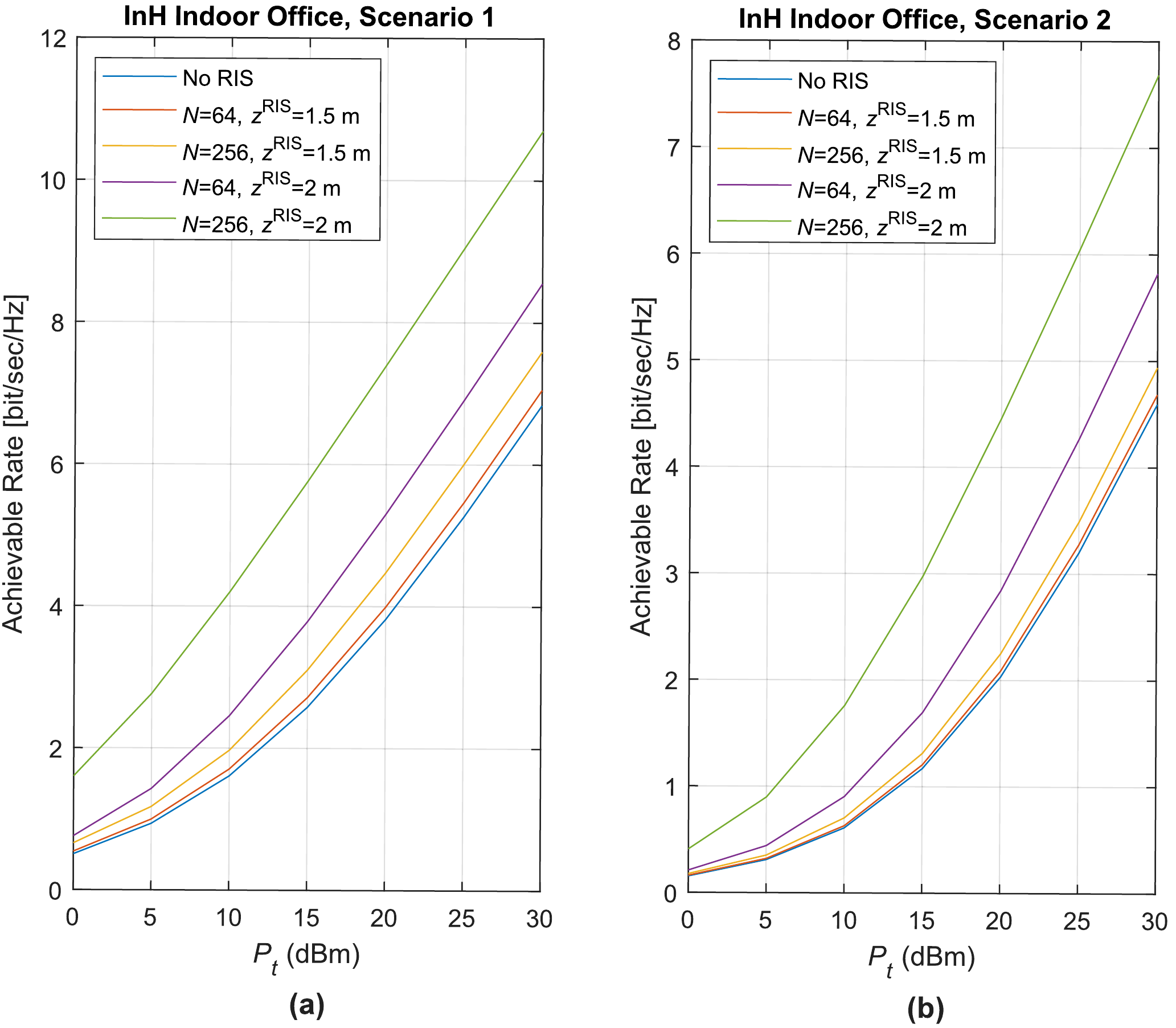}
		\vspace*{-0.3cm}\caption{Achievable rates of RIS-assisted systems for indoors under (a) Scenario
			1 and (b) Scenario 2.}\vspace*{-0.5cm}
		\label{fig:Sim1}
	\end{center}
\end{figure}

In the following, basic numerical results are presented using the proposed channel model for RIS-empowered communication via the open-source \textit{SimRIS Channel Simulator} MATLAB package. We consider the operating frequency of $28$ GHz, while the considered system parameters are also valid for $73$ GHz. The noise power is assumed to be $-100$ dBm for the calculation of ergodic achievable rate: $ R=\mathrm{E}\left\lbrace \log_2(1+\rho)\right\rbrace $, where $ \rho $ is the instantaneous received signal-to-noise ratio.

In Figs. \ref{fig:Sim1}(a) and (b), we evaluate the achievable rate $(R)$ of a communication system with and without an RIS operating in indoor environments. Here $R$ is defined as $R=\mathrm{E}\left\lbrace \log_2(1+\rho)\right\rbrace $ [bits/s/Hz]. In Fig. \ref{fig:Sim1}(a), we consider Scenario 1 where the RIS is mounted on the side wall and the coordinates of the Tx, the Rx, and the RIS are respectively given as $(0,25,2)$, $(38,48,1)$, $(40,50,z^{\text{RIS}})$. Similarly, in Fig. \ref{fig:Sim1}(b), we consider Scenario 2 with the Tx, the Rx, and the RIS coordinates given by $(0,25,2)$, $(70,35,1)$, and $(70,30,z^{\text{RIS}})$, respectively.  Here, we consider two different RIS placements: RIS mounted at a moderate height of $z^{\text{RIS}}=1.5$ m (low LOS probability) and at a higher height of $z^{\text{RIS}}=2$ m ($ 100\% $ LOS probability). For these given parameters, we have $d_{\text{RIS-R}} \in \left\lbrace 2.87,3 \right\rbrace $ m and $d_{\text{RIS-R}} \in \left\lbrace 7.09,7.14 \right\rbrace $ m for Scenarios 1 and 2, respectively, which are valid assumptions to have a pure LOS link between the RIS and the Rx.  As seen from  Figs. \ref{fig:Sim1}(a) and (b), for the case of $z^{\text{RIS}}=1.5$ m, since the LOS probability is relatively low for the Tx-RIS link, which has a LOS distance of $47.1$ m, the RIS provides only a minor improvement in the received SNR. The reason of this behavior can be explained by the relatively higher attenuation of the RIS-assisted channel compared to the channel between Tx-Rx. However, a major improvement is observed for the case of $z^{\text{RIS}}=2$ m, which assumes a LOS-dominated Tx-RIS link. From the given results of Fig. \ref{fig:Sim1}, we conclude that the RIS can be used as an effective tool to boost the achievable rate in indoor environments when both the Tx-RIS and RIS-Rx links are LOS dominated.

\begin{figure}[!t]
	\begin{center}
		\includegraphics[width=0.9\columnwidth]{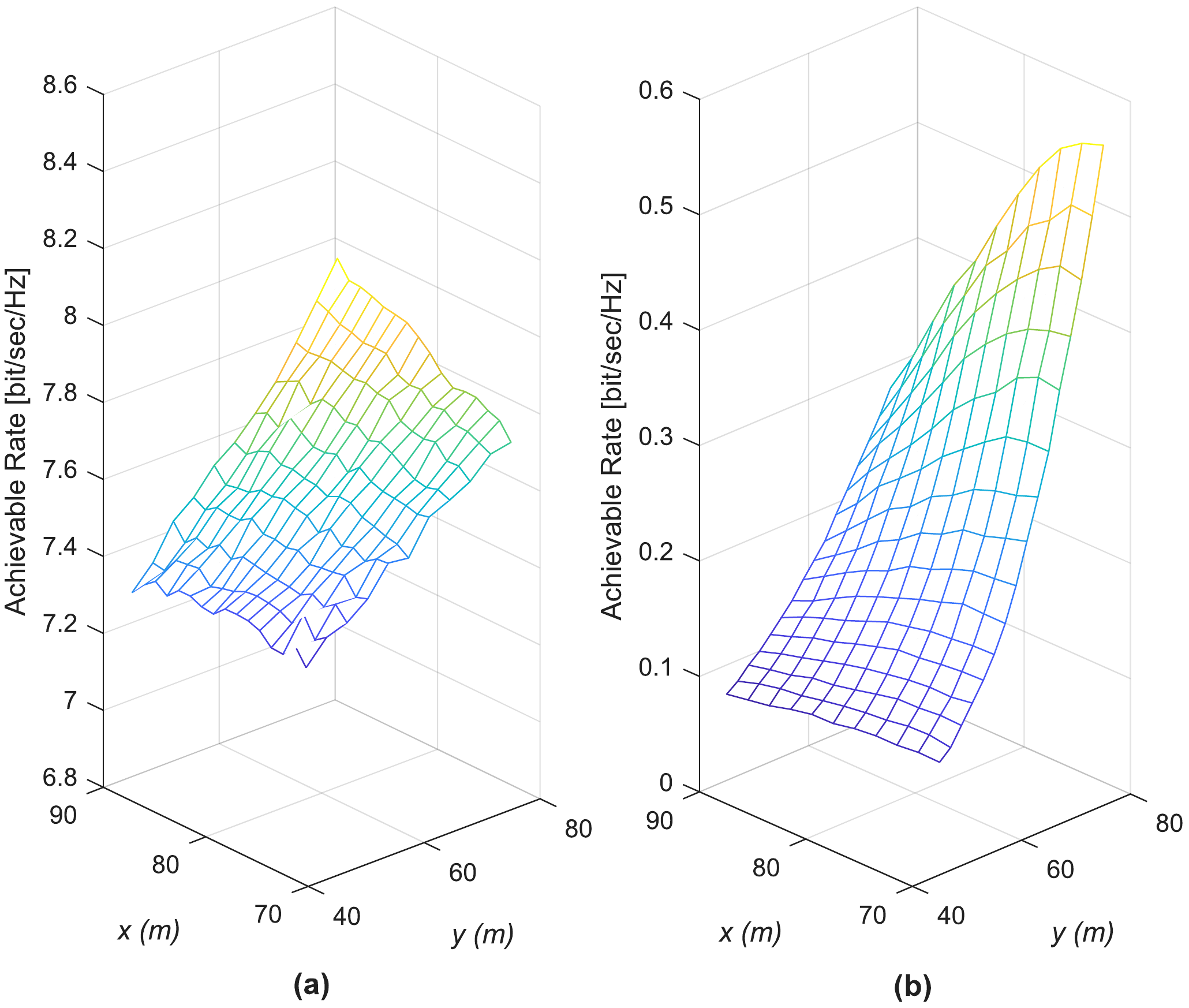}
		\vspace*{-0.3cm}\caption{Achievable rates of the an RIS-assisted system under varying Rx positions in an outdoor environment in the presence (a) and absence (b) of the direct link between the Tx-Rx.}\vspace*{-0.6cm}
		\label{fig:Sim3}
	\end{center}
\end{figure}

In Figs. \ref{fig:Sim3}(a) and \ref{fig:Sim3}(b), the effect of varying Rx positions through the $ x $ and $ y $-axis on the achievable rate of an RIS-assisted system at $28$ GHz is examined for a UMi Street Canyon outdoor environment. 
Here, we consider Scenario 1 where the RIS is mounted on the $xz$ plane and the coordinates of the Tx, the Rx, and the RIS are respectively given as $(0,25,20)$, $(x^{\text{Rx}},y^{\text{Rx}},1)$, $(70,85,10)$. In Fig. \ref{fig:Sim3}(a), the direct link between the Tx-Rx is available as well as the RIS-assisted link for transmission, while it is assumed that the direct link between Tx-Rx is blocked in Fig. \ref{fig:Sim3}(b). In Fig. \ref{fig:Sim3}(a), we observe that the highest achievable rate is obtained when the Rx is close to the Tx, since the channel between Tx-Rx is more dominant than the RIS-assisted link in terms of achievable rate. Furthermore, if the Rx moves to an area far from the RIS on the $ x $ and $y$-axis, the effect of the RIS will drastically diminish as shown in Fig. \ref{fig:Sim3}(b). We observe that the most decisive performance parameter is the separation between the RIS-Rx when the direct link is blocked between the Tx-Rx.

\section{Conclusions}
This paper has been a first step towards physical channel modeling for emerging RIS-empowered networks and aimed to fill an important gap in the open literature by providing a open-source and widely applicable physical channel model for mmWaves. Our \textit{SimRIS Channel Simulator} package can be used effectively in channel modeling of RIS-based systems with optional operating frequency, terminal locations, number of RIS elements, and environments. Since the fundamental aim of this study is to pave the way for an accurate channel model which is compatible with real-world experiments, our future endeavor will be to extend this work by considering various conditions such as delay spread and fast fading for MIMO channel models to introduce a holistic RIS-assisted channel model by obtaining experimental results.. We invite all researchers working in this field to help the testing and development of \textit{SimRIS}.

\bibliographystyle{IEEEtran}
\bibliography{bib_2020}

\end{document}